# Using LSTM Encoder-Decoder Algorithm for Detecting Anomalous ADS-B Messages

Edan Habler            Asaf Shabtai

Department of Software & Information Systems Engineering
Ben-Gurion University of the Negev
Beer-Sheva 8410501, Israel
habler@post.bgu.ac.il, shabtaia@bgu.ac.il

*Abstract*—Although the ADS-B system is going to play a major role in the safe navigation of airplanes and air traffic control (ATC) management, it is also well known for its lack of security mechanisms. Previous research has proposed various methods for improving the security of the ADS-B system and mitigating associated risks. However, these solutions typically require the use of additional participating nodes (or sensors) (e.g., to verify the location of the airplane by analyzing the physical signal) or modification of the current protocol architecture (e.g., adding encryption or authentication mechanisms.) Due to the regulation process regarding avionic systems and the fact that the ADS-B system is already deployed in most airplanes, applying such modifications to the current protocol at this stage is impractical. In this paper we propose an alternative security solution for detecting anomalous ADS-B messages aimed at the detection of spoofed or manipulated ADS- B messages sent by an attacker or compromised airplane. The proposed approach utilizes an LSTM encoder-decoder algorithm for modeling flight routes by analyzing sequences of legitimate ADS-B messages. Using these models, aircraft can autonomously evaluate received ADS-B messages and identify deviations from the legitimate flight path (i.e., anomalies). We examined our approach on six different flight route datasets to which we injected different types of anomalies. Using our approach we were able to detect all of the injected attacks with an average false alarm rate of 4.3% for all of datasets.

## I. INTRODUCTION

Over the last decade, there has been a significant increase in the number of flight movements around the world, with an average of approximately 100,000 registered flight movements per day, estimated by the International Air Transport Association (IATA) in 2015.[1] Due to the growing need for civilian flights and the adoption of unmanned aerial vehicles (UAC), the number of registered flight movements around the world will undoubtedly continue to increase. In fact, according to IATA forecasts, this number is predicted to soar and will likely double by 2035.[2]

In order to provide safe navigation and reduce the cost of air traffic control (ATC), the aviation community has been moving from uncooperative and independent air traffic surveillance, such as primary surveillance radar (PSR) or secondary surveillance radar (SSR), to cooperative and dependent air traffic surveillance (CDS), such as ADS-B.

Automatic dependent surveillance-broadcast (ADS-B) [1] is a modern implementation of SSR certified by the International Civil Aviation Organization (ICAO) and the Federal Aviation Administration (FAA) which is expected to play a major role in aviation in the future. The ADS-B system provides the ability to continuously and precisely localize aircraft movements in dense air space. An aircraft equipped with an ADS-B transponder (transmitter-responder) is capable of deriving its position from the navigation satellite system, and then broadcasts the aircraft's flight number, speed, position, and altitude at an average rate of 4.2 messages per second.

Unlike issues of cost and accuracy, which were major considerations in the development of ADS-B, security was pushed to the sidelines. This resulted in a widely used technology with highly compromised security, particularly in terms of the protocol mechanism, as follows:

**No message authentication and encryption:** messages are broadcast as plain text without an authentication code or digital signature and therefore can be replayed, manipulated, or forged.

**No aircraft authentication:** authorized aircraft or ATC stations don't have to authenticate before transmitting; thus, there is no way to distinguish between authorized and unauthorized entities. As a result, an unauthorized entity can inject messages or tamper with an authorized entity's reports.

Previous research has demonstrated that it is relatively easy to compromise the security of ADS-B with off-the-shelf hardware and software [2][3]. The ability to exploit the ADS-B system endangers billions of passengers every year, and therefore there have been attempts by academia and industry to develop solutions that address the lack of security.

Past research suggested the use of encryption [4], aircraft authentication via challenge-response [5], and message authentication [3][6], in order to provide secured message broadcast and prevent eavesdropping. Besides securing broadcast communication, additional approaches focused on verifying velocity and location reports via additional sensors or nodes. However, most of those solutions require modifications to the architecture in order to enable key exchange or establish trust between entities. Since the FAA has mandated the use of ADS-B for all aircraft movements within the US airspace by 2020, a requirement that already exists for some aircraft in Europe, and due to the strict regulation process regarding the implementation of avionic systems, applying modifications to the current protocol at this stage is impractical (note that the ADS-B protocol design and development began in the early 1990s).

---
[1] http://www.iata.org/about/Documents/iata-annual-review-2016.pdf
[2] http://www.iata.org/pressroom/pr/Pages/2016-10-18-02.aspx

In this study we propose an alternative security solution for detecting anomalous ADS-B messages; specifically, our approach is aimed at detecting spoofed or manipulated ADS-B messages sent by an attacker or compromised airplane. The proposed approach does not require any modification or additional participating nodes and/or sensors, and enables aircraft to detect anomalies in the dense air space autonomously. Our approach is designed to address message spoofing by observing a sequence of messages and estimating its credibility. Since flights between airports usually take place via similar routes, we use and train an LSTM (long short-term memory) encoder-decoder model based on previous (legitimate) flights for a given route. Using such a model, each aircraft can independently evaluate received ADS-B messages and identify deviations from the legitimate flight path.

We examined our approach using six datasets, each dataset contains flight information for a selected route. In our experiment we injected different types of anomalies (erroneous data) into this data and demonstrated that our approach was able to detect all of the injected attacks with an average of 4.303% false alarm rate. In addition, we measured the alarm delay as the number of messages sent from the moment the attack started until detection.

The contributions of our paper are as follows. First, to the best of our knowledge, we are the first to utilize machine learning techniques to secure the ADS-B protocol; specifically, we show that sequences of ADS-B messages can be modeled by using the LSTM encoder-decoder algorithm. Second, we show that the LSTM encoder-decoder model can be used to amplify anomalies and thus facilitates the detection of anomalous messages. Third, our proposed approach can overcome ADS-B shortcomings, particularly in the case of spoofed/fake messages, using standalone solutions that do not require architecture changes and can be applied by each aircraft independently. Finally, the proposed model is adaptive and flexible, so it can be trained and applied to new routes.

The rest of the paper is organized as follows: Section II provides an overview of the ADS-B protocol and its security risks, and Section III presents prior studies that proposed security solutions for the ADS-B protocol. Section IV contains a description of our proposed approach for detecting anomalous ADS-B messages and our feature extraction process. Section V describes our evaluation and discusses the results. Finally, Section VI concludes the paper and mentions future work directions.

## II. OVERVIEW OF ADS-B PROTOCOL AND RELATED RISKS

### A. Protocol Overview

Automatic dependent surveillance-broadcast (ADS-B) is a satellite-based 'radar-like' system that automatically, independently, and continuously derives the aircraft's position from the global navigation satellite system (e.g., GPS, GLONASS, and Galileo) and broadcasts the data to nearby aircraft and ground stations. ADS-B was developed in order to improve air traffic control and was rolled out as a replacement to traditional primary/secondary radar. Providing improved accuracy and greater coverage in both radar and non-radar environments (e.g., mountain areas and oceans), ADS-B is designed to prevent collisions and improve utilization and throughput of aircraft in dense airspace.

The system includes two subsystems: ADS-B Out and ADS-B In. The ADS-B In subsystem enables aircraft to receive broadcast messages of other nearby aircraft. The ADS-B Out subsystem enables aircraft to continually broadcast messages. The system enables an aircraft to broadcast unencrypted messages that provide the position of the aircraft, its velocity, and its altitude, as well as additional information, using the ADS-B Out subsystem. The transmitted messages are processed by nearby aircraft and ATC stations on the ground using the ADS-B In subsystem (illustrated in Figure 1).

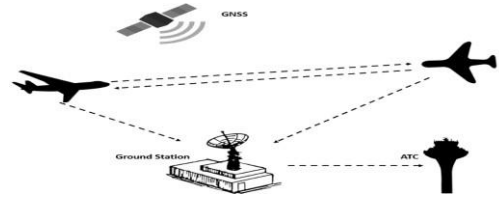

**Fig. 1.** An illustration of the ADS-B system. Position is provided by the GNSS, processed by the aircraft, and broadcast by the ADS-B Out subsystem. Ground stations and nearby aircraft receive these messages via the ADS-B In subsystem. Ground stations then transmit the aircraft's data to ATC.

### B. Risk Overview

The ADS-B system is lacking basic security mechanisms such as authentication, message integrity, and encryption. In light of the need for real-time information, these security gaps make the application of the protocol in the crowded skies risky, exposing aircraft to the following types of attacks:

- **Eavesdropping**: The lack of message encryption and insecure broadcast transmissions makes eavesdropping over the medium by both adversaries (e.g., potential terrorist groups) and non-adversaries (e.g., commercial Internet websites) easy, allowing outsiders to track air traffic. McCallie, *et al.* [7] describes "Aircraft Reconnaissance" as an eavesdropping attack that may target specific aircraft or seek to gain information about the air traffic.

- **DoS:** Denial of service attacks can have a significant impact on real-time systems like ADS-B. For example, a jamming attack, in which a single participant is prevented from sending or receiving messages by an attacker. Wilhelm, *et al.* [8] demonstrated feasible real-time jamming in wireless networks, thus raising concern in aviation networks.

- **Spoofing via message injection/deletion:** Since there are no challenge-response mechanisms in the ADS-B system, neither entities (sender and receiver) are authenticated. Thus, an attacker can broadcast forged messages using low cost commercial off-the-shelf (COTS) software, impersonate an authorized entity, or even inject ghost aircraft information, as illustrated in [9]. Furthermore, message modification via different approaches, namely overshadowing, bit-flipping, and message deletion, using destructive and constructive interference are discussed by Strohmeier in [10].

*C. Adversary Model*

Understanding the adversary model is essential in order to estimate an attacker's capabilities of performing the attacks mentioned above. We distinguish between the following two kinds of attackers:

- **External attacker** – an external attacker is an adversary that can execute simple attacks using COTS transponders. In order to transmit signals, one does not have to authenticate or belong to a specific airline. Therefore, while standing on the ground, an external attacker could receive and transmit signals, and perform DoS, eavesdropping, and spoofing attacks, however as suggested by Strohmeier *et al.* [11], there are several approaches to detect an attacker in this situation (e.g., measuring the Pearson correlation coefficient between the claimed aircraft's position and the received signal strength). More complicated attacks performed via UAVs or drones will be much harder to detect via signal analysis.
- **Internal attacker** – an internal attacker is an adversary that has achieved access to the system and affects its behavior (e.g., an ATC crew member or aircraft maintenance worker). An internal attacker can manipulate the data processing phase or disrupt the system modules.

III. RELATED WORK

Previous publications investigated the security challenges associated with the ADS-B system and proposed various methods and solutions for protecting the system. The main ideas in previous research include encryption, physical layer analysis, and multilateration technique. The following topics provides an overview of relevant methods and work.

*A. Encryption*

Cryptographic measures have been tested for securing communication in wireless networks. Strohmeier *et al.* [10] discussed the question of whether the current implementation of ADS-B can be encrypted. In addition, Finke *et al.* [4] introduced a number of encryption schemes. However, the worldwide deployment of the ADS-B system makes the encryption key management a challenge. Costin *et al.* [3] and Feng *et al.* [6] suggested PKI (public key infrastructure) solutions based on transmitting signatures and the Elliptic Curve cryptography, respectively. Another solution discussed in the literature is the use of the retroactive key publication technique, such as the use of μTESLA protocol [10]; this however requires modifications to the current mechanism of the protocol by adding a new message type for key publishing.

*B. Physical Layer Analysis and Doppler Effect*

One of the most dangerous types of attack is spoofed message injection. Strohmeier *et al.* [11] proposed an intrusion detection system, based on physical layer information and a single receiver, in order to detect such attacks on critical air traffic infrastructures without additional cooperation by the aircraft. Another solution suggested by Ghose *et al.* [12] is to verify the velocity and position of the aircraft by exploiting the short coherence time of the wireless channel and the Doppler spread phenomenon. A method presented by Schäfer [13] is based on verification of the motion using the Doppler effect. Both options rely upon the participation of ground stations or other entities.

*C. Multilateration and Group Verification*

In the last decade, signal analysis has been successfully employed in the fields of wireless communication. One popular form of cooperative independent surveillance that has been used in the military and civil applications is multilateration (MLAT). MLAT is a navigation technique based on the measurement of the time difference of arrival (TDOA) between at least two stations at known locations. A method to provide a means for back-up ADS-B communication based on MLAT was provided by Smith [16]. Additional work based on MLAT suggested by Schäfer *et al.* [14] showed that it is possible to verify a 3D route with a group of four verifiers. In addition, Strohmeier *et al.* [15] suggested a method of continuous location verification by computing the differences in the expected TDOA between at least two sensors.

IV. PROPOSED MODEL

*A. Motivation*

Detecting anomalies using standard approaches of predictive models, especially when detecting anomalies in a time series, is a challenging task, since the context of the current sample and its past may influence its value.

We therefore opt to use an LSTM encoder-decoder algorithm in order to profile flight routes and detect anomalies. The use of applying machine learning (specifically deep learning models) does not require modifications to the current architecture of the ADS-B system or additional participating nodes. This allows the aircraft to autonomously and independently analyze ADS-B messages for anomaly detection.

We define an ADS-B *window* of size $n$ as a sequence of $n$ consecutive ADS-B messages. A *malicious window* is defined as a *window* which includes at least one spoofed ADS-B message. The window containing all messages of a flight from the $i$-th message to the $i+n$ message is denoted by $W[i,n] = \{x^{(i)}, x^{(i+1)}, ..., x^{(i+n)}\}$. Each entry $x^{(j)}$ is a vector consisting of features extracted for message $j$ during the flight.

The LSTM encoder-decoder algorithm is utilized for detecting anomalous (malicious) windows. This is done by training an encoder-decoder model for a route from takeoff point A to landing point B. During the training phase we fit the model to reconstruct normal (benign) *windows* of flights from point A to point B. For each tested *window* we first use LSTM in order to encode the sequence of ADS-B messages

(where each message is represented by the vector of features) to a fixed dimension vector (i.e., sequence to sequence model). Then we use a decoder based on LSTM to decode and reconstruct the tested *window*. When the model reconstructs an anomalous *window* it may not reconstruct the sequence well and will therefore amplify the reconstruction error.

### B. Feature Extraction

In order to be able to differentiate between normal and anomalous windows, the extraction of meaningful features that will provide the context of the flight is required.

First, we extract the aircraft's speed, geolocation (latitude / longitude), altitude, and heading from each message. In order to provide contextual flight-progress, we also extract representative features for each flight. This is done by computing the average path of a route (using previous legitimate flight records) and extract four major geolocation points for each source and destination (illustrated in Figure 2).

Afterwards, we measure the distance between each point in the route (latitude, longitude, as received by the messages) and the aforementioned major points (see Figure 2) using the inverse method of Vincenty's distance formulae [17].

### C. Model Description

#### 1) Training the LSTM encoder-decoder model

As presented by Malhotra *et al.* in [18], we train an LSTM encoder-decoder model to reconstruct *windows* of benign sequences with minimal error; i.e., the model attempts to output the same input sequence of vectors.

This is achieved by using an LSTM encoder that learns from fixed length sequences of messages (each message is represented by the vector of features) by optimizing the hidden layer ($H_D$). The LSTM decoder reconstructs the *window* using the current hidden state of the decoder ($H_D$) and the values predicted in the previous message (see Figure 3).

#### 2) Applying the LSTM encoder-decoder model

Since during the training phase the model is trained to reconstruct a legitimate sequence of messages (i.e., a *window*), we expect both of the model's inputs and outputs to look alike. In contrast, when we apply the model on a malicious *window* (i.e., containing spoofed messages), we expect the model to fail at reconstructing it, and therefore input vectors (input *window*) and output vectors (predicted *window*) will differ significantly.

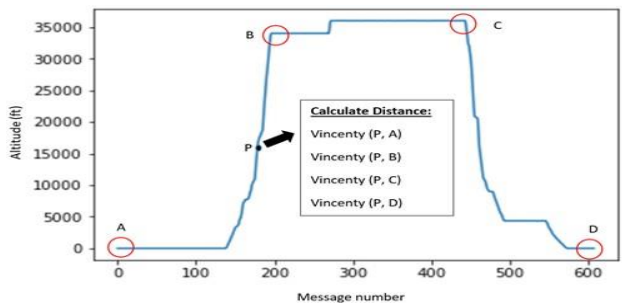

**Fig. 2.** An average flight from the London dataset. For each point P in the flight, the Vincenty distance from A, B, C, and D is calculated.

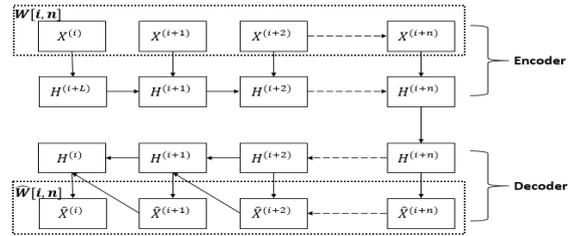

**Fig. 3.** Illustration of the LSTM-based encoder-decoder. The steps for obtaining the prediction of *window* $\widehat{W}[i,L]$ from input window $W[i,L]$ are as follows: (1) the encoder encodes the input vectors into a fixed sized vector, and (2) the decoder decodes the fixed sized vector in an attempt to reconstruct the original *window*.

After predicting the output window $\widehat{W}[i,L]$ corresponding to the target input window $W[i,L]$, we obtain the reconstruction error of each vector representing an ADS-B message using the Cosine similarity (see Equation 1). The overall anomaly score of the input window is computed according to Equation 2.

$$CosSimilarity(x,\hat{x}) : \frac{\sum_{i=1}^{n} x_i \hat{x}_i}{\sqrt{\sum_{i=1}^{n} x_i^2} \sqrt{\sum_{i=1}^{n} \hat{x}_i^2}} \quad (1)$$

$$Anomaly(W[i.L]) = \sum_{j=i}^{i+L}(1 - CosSimilarity(x^{(j)}, \hat{x}^{(j)}) \quad (2)$$

## V. EXPERIMENTS

We conducted a set of experiments in order to evaluate our proposed approach, and more specifically, the ability of our approach to model an arbitrarily chosen route and use this model to deduce whether a given flight (or a segment of the flight) is benign or an anomaly.

### A. Dataset

**Collected dataset**: We used a large-scale dataset from the online flight tracking network, FlightRadar24,[3] for our evaluation. FlightRadar24 provides access to data collected from thousands of ground stations. The extracted datasets are presented in Table I.

**Injected anomalies**. In order to evaluate the performance of the learned model, we injected three types of anomalies (a segment of 70 sequential messages, from message 180 to message 250) into the flights included in the test sets:

*Random noise (RND)* – anomalies are generated by adding random noise. We multiplied the original values of the message attributes of the ADS-B messages with a randomly generated floating number between 0 and 2.

*Different route (ROUTE)* – anomalies are generated by replacing a segment of the ADS-B messages of the tested flight with a segment of messages from a different (legitimate) route. In our evaluation, we replaced a segment from the flights in our datasets with segment from the flight between Suvarnabhumi Airport, Thailand and Tashkent International Airport, Uzbekistan (the Thailand dataset).

*Gradual drift (DRIFT)* – anomalies are generated as a gradual drift in the altitude feature. This is done by modifying the -

---
[3] http://www.flightradar24.com

TABLE I. DESCRIPTION OF DATASETS

| Dataset Name | Description | Number of flights | Approx. flight duration | Dates of flights |
|---|---|---|---|---|
| London | Flights between Ben Gurion Airport, Israel and Luton Airport, London | 80 | 5 hours | May 15, 2017 – July 1, 2017 |
| Milano | Flights between Ben Gurion Airport, Israel and Malpensa Airport, Italy | 65 | 3.75 hours | March 7, 2017 – July 27, 2017 |
| Moscow | Flights between Moscow Sheremetyevo Airport, Russia and Heathrow Airport, London | 54 | 4 hours | May 3, 2017 – July 7, 2017 |
| Washington | Flights between San Francisco Airport, United States and Washington Dulles Airport, United States | 70 | 4.3 hours | April 10, 2017 – July 9, 2017 |
| Paris | Flights between Kiev Boryspil Airport, Ukraine and Paris Charles de Gaulle Airport, France | 68 | 3 hours | April 3, 2017 – July 9, 2017 |
| Las Vegas | Flights between Benito Juarez Airport, Mexico and Las Vegas McCarran Airport, United States | 80 | 3.3 hours | February 20, 2017 – July 7, 2017 |
| Thailand *(used for generating anomalies)* | A flight between Suvarnabhumi Airport, Thailand and Tashkent International Airport, Uzbekistan | 1 | 6 hours | July 6, 2017 |

altitude of a segment of messages by continuously raising/lowering the altitude by an increasing multiplier of 400 feet (i.e., for the first message in the anomalous segment the altitude will be increased/decreased by 400 feet, the second message will be increased/decreased by 800 feet, and so on). In our evaluation we generated two types of *gradual drifts* by lowering the altitude value (denoted as *SHIFT Down*) and raising the altitude value (denoted as *SHIFT Up*).

By selecting and evaluating these types of anomalies we are able to represent two types of attackers. The first, is a naïve adversary (RND and ROUTE anomalies) with the goal of adding observable noise to the air-space view in order to reduce the credibility of the ADS-B system and disrupt the traffic management. The second adversary is less aggressive and more sophisticated that attempts to influence the air-space view by adding reasonable (less observable) gradually-drifted messages (in location or altitude) which may result in a collision in air.

### B. Evaluation Approach and Configuration

The experiments were conducted using the 10-fold cross-validation approach as follows. We divided the flights of each dataset into 10 folds, each containing an equal number of flights. For each fold $i$ and dataset $DS$ (London, Milano, Moscow, Washington, Paris, and Las Vegas) the training set includes all of the flights in $DS$, excluding the flights of the $i$-th fold (denoted by $Train_i^{(DS)}$); the flights of the $i$-th fold are used for testing (denoted by $Test_i^{(DS)}$). The $Test_i^{(DS)}$ dataset was duplicated four times; for each copy, a set of malicious *windows* were injected (as mentioned from message 180 to message 250) according to the four types of anomalies (one type of anomaly for each copy). We denote these datasets by: $Test_{i,RND}^{(DS)}, Test_{i,ROUTE}^{(DS)}, Test_{i,SHIFT\ UP}^{(DS)}, Test_{i,SHIFT\ DOWN}^{(DS)}$.

In our experiments the *window* size (i.e., the size of the sequence input to the LSTM encoder-decoder model) was set at $L=15$. In addition, in order to evaluate the model derived from the training set, we defined a *window* that contains 15 messages as a malicious *window* if it contains at least one spoofed message. In order to set the threshold value for an anomalous *window*, we performed 5-fold cross-validation evaluation on $Train_i^{(DS)}$. Since the $Train_i^{(DS)}$ dataset includes benign flights only, we obtained the anomaly scores (Equation 2) and defined the value that exceeds 95% of the errors as the threshold value for the testing phase of: $Test_{i,RND}^{(DS)}, Test_{i,ROUTE}^{(DS)}, Test_{i,SHIFT\ UP}^{(DS)}, Test_{i,SHIFT\ DOWN}^{(DS)}$.

To assess the performance of the models, we examined the corresponding false positive rate (FPR), true positive rate (TPR), and the alarm delay of the model (measured as the number of messages from the beginning of the attack until a malicious *window* is detected).

### C. Results

Figure 4 contains a graphical representation of the anomaly score for each type of attack for a single representative flight, randomly chosen from the London dataset. It illustrates the increase in the anomaly score as the anomaly becomes more significant. This is because the evaluated input *window* contains an increasing number of anomalous message. The Gradual Drift anomaly (SHIFT DOWN) of the same selected flight is also visualized on top of a geographical map in Figure 5. Each icon indicate a *window* of ADS-B messages where the actual location is set according to the location of the last ADS-B message in the *window*.

The results of the experiments are presented in Table II. Table II shows the average and standard deviation of the FPR, TPR, and alarm delay time for each type of attack. We can infer from the results that the proposed model can efficiently predict an ongoing anomaly, while the alarm delay time changes according to the attack's aggressiveness. As can be seen, attacks of type *RND* and *ROUTE* were detected almost immediately. This is due to the fact that they affected more than one attribute of the ADS-B message. On the other hand, the *SHIFT Down* and *SHIFT Up* type of attacks affected only one attribute (altitude), therefore the delay time was longer.

In an attempt to reduce the rate of false alarms, we examined the results using a collective (aggregative) anomaly by raising an alert only when a sequence of $t$ malicious *windows* were detected. We examined the detection rate and false alarm rate for $t=5$, 10, and 15. The results are presented in Table III which shows the false alarm rate for each dataset and for different values of $t$ averaged for all folds and attacks.

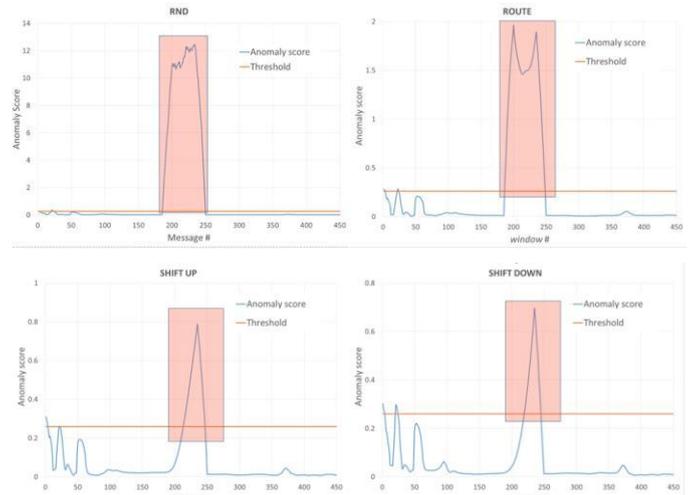

**Fig. 4.** Example of a flight with injected anomalies. The malicious *windows* are represented by red rectangles.

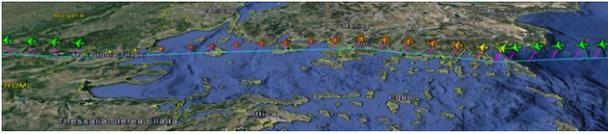

**Fig. 5.** SHIFT DOWN visualization of a selected flight. Each icon indicates the location of the aircraft and represents a window of ADS-B messages. The size of the icon indicates the reported altitude of the aircraft and the color of the icon indicates the anomaly level derived by the relevant model (Red being an anomalous window and Green benign window).

TABLE II. EXPERIMENTS RESULTS

|  | RND | ROUTE | SHIFT UP | SHIFT DOWN |
|---|---|---|---|---|
| **London Dataset** | | | | |
| **Alarm delay** | 1.0±0.0 | 1.0±0.0 | 26.91±2.15 | 26.49±2.42 |
| **FPR** | 5.56±1.30% | 5.48±1.41% | 5.34±1.38% | 5.34±1.30% |
| **TPR** | 98.43±1.30% | 97.13±0.23% | 36.87±4.10% | 45.12±3.99% |
| **Milano Dataset** | | | | |
| **Alarm delay** | 1.0±0.0 | 1.0±0.0 | 33.17±1.17 | 29.34±1.45 |
| **FPR** | 5.75±2.04% | 5.75±2.04% | 5.47±2.05% | 5.47±2.05% |
| **TPR** | 98.44±0.00% | 98.44±0.00% | 43.14±2.20% | 51.92±2.43% |
| **Moscow Dataset** | | | | |
| **Alarm delay** | 1.0±0.0 | 1.0±0.0 | 33.333±2.04 | 28.816±2.92 |
| **FPR** | 4.2±1.46% | 4.2±1.46% | 3.87±1.45% | 3.87±1.46% |
| **TPR** | 98.82±0.06% | 98.82±0.06% | 59.04±2.30% | 65.47±2.23% |
| **Washington Dataset** | | | | |
| **Alarm delay** | 1.0±0.0 | 1.0±0.0 | 35.65±2.28 | 30.97±2.70 |
| **FPR** | 5.14±1.15% | 5.1±1.15% | 4.86±1.15% | 4.87±1.00% |
| **TPR** | 98.80±0.00% | 98.80±0.00% | 55.52±3.12% | 63.06±3.67% |
| **Paris Dataset** | | | | |
| **Alarm delay** | 1.0±0.0 | 1.0±0.0 | 37.44±5.30 | 31.74±2.14 |
| **FPR** | 5.39±2.78% | 5.39±.78% | 4.96±2.71% | 5.15±2.80% |
| **TPR** | 98.84±0.00% | 98.84±0.00% | 51.93±6.96% | 61.37±2.28% |
| **Las Vegas Dataset** | | | | |
| **Alarm delay** | 1.0±0.0 | 1.0±0.0 | 35.45±2.72 | 34.50±3.86 |
| **FPR** | 5.52±0.92% | 5.52±0.92% | 5.20±0.91% | 5.20±0.91% |
| **TPR** | 98.80±0.00% | 98.80±0.00% | 49.49±3.37% | 48.95±5.74% |

TABLE III. AVERAGE FALSE ALARM RATE

| Dataset Name | $t=5$ | $t=10$ | $t=15$ |
|---|---|---|---|
| London | 4.70 | 1.26 | 0.55 |
| Milano | 8.70 | 2.21 | 0.57 |
| Moscow | 2.51 | 0.75 | 0.53 |
| Washington | 4.20 | 2.28 | 1.28 |
| Paris | 2.35 | 0.35 | 0.05 |
| Las Vegas | 0.20 | 0.11 | 0.05 |

As can be observed, the lowest false alarm rate was attained for t=15. Note that in all cases the true attack was detected; that is, the detection rate is 1.0.

## VI. CONCLUSION AND FUTURE WORK

In this study, we proposed a flexible and adaptive time series anomaly detection scheme for false data injected into ADS-B messages, based on an LSTM encoder-decoder model. We validated our model using six FlightRadar24 datasets, injected with four types of false data. We detected each injection type with a low false alarm rate. Based on this, we can deduce that it is possible to detect anomalies or estimate the legitimacy of messages without changing the ADS-B protocol or its underlying architecture. In future work we plan to evaluate our approach using anomalies representing more sophisticated attacks and compare the results of the model after tuning various hyper-parameters and adding more features. In addition, we plan to model the airspace state in a certain geo-location in order to provide more contextual information. This will enable the detection of replay attacks and DoS attacks.